\documentclass[prb,twocolumn,showpacs]{revtex4}
\usepackage{graphicx}
\usepackage{times}
\usepackage{dcolumn}
\usepackage{natbib}
\begin{document}
\title{Electronic states in 1/1 Cd$_6$Yb and 1/1 Cd$_6$Ca: 
Relativistic, correlation, and structural effects} 
\author{E.\ S.\ Zijlstra}
\email[]{ezijlstr@brocku.ca}
\author{S.\ K.\ Bose}
\affiliation{Department of Physics, Brock University, St.\ Catharines, Ontario
L2S 3A1, Canada}
\author{R.\ Tamura}
\affiliation{Department of Materials Science and Technology, Tokyo
University of Science, Noda, Chiba 278-8510, Japan}
\date{\today}
\begin{abstract}
The electronic structure of the rational approximants 1/1 Cd$_6$Yb and 1/1 
Cd$_6$Ca to the stable icosahedral CdYb and CdCa quasicrystals is studied by 
the full-potential linear augmented plane wave method.
A comparison is made between several structural models.
We show that the (relativistic) spin-orbit (SO) interaction and electronic 
correlations that are not described by the usual local density approximation, 
are essential for an accurate description of the electronic structure.
In particular, we show that the SO interaction is responsible for a splitting 
of the Cd-4\textit{d} and Yb-4\textit{f} peaks, and that the experimental peak 
positions can be reproduced by including a Hubbard $U$ term in the Hamiltonian 
[$U$(Cd) $= 5.6$ eV, $U$(Yb) $= 3.1$ eV]. 
Our results show very good agreement with a photo-emission (PE) spectrum of 
1/1 Cd$_6$Yb [R.\ Tamura, Y.\ Murao, S.\ Takeuchi, T.\ Kiss, T.\ Yokoya, and 
S.\ Shin, Phys.\ Rev.\ B \textbf{65}, 224207 (2002)] and a $350$ eV PE 
spectrum of 1/1 Cd$_6$Ca, which we present in this paper.
Without the relativistic and correlation effects even a qualitative agreement
with the PE spectra cannot be achieved.
\end{abstract}
\maketitle

The recently discovered binary icosahedral quasicrystals (QCs) 
\textit{i}-CdYb \cite{Guo00cdyb} and \textit{i}-CdCa \cite{Guo00cdca} have
received considerable attention.
Their electronic structures have been calculated and analyzed 
\cite{Ishii01} from the viewpoint of their stability and differences with
respect to ternary Al-based QCs.
The purpose of this communication is to point out that two elements
ignored in previously published calculations, namely, the spin-orbit (SO)
coupling and correlation effects, are of crucial importance for an accurate
description of the electronic structure of these QCs.
In addition, this communication addresses the issue of the stability of 
various structural models proposed for the 1/1 approximants of these QCs.

The binary icosahedral QCs \textit{i}-CdYb and \textit{i}-CdCa were discovered
\cite{Guo00cdyb,Guo00cdca} in 2000.
Prior to this discovery the most studied QCs were the ternary alloys
\textit{i}-AlCuFe, \textit{i}-AlPdMn, \textit{i}-AlPdRe, \textit{d}-AlNiCo, 
and \textit{d}-AlCuCo.
It has been anticipated \cite{Guo00cdyb} that binary QCs simplify the 
analysis of experimental properties.
The study of \textit{i}-CdYb and \textit{i}-CdCa has indeed shed light on 
properties that are arguably still not well understood in the
above-mentioned ternary QCs, such as the negative temperature coefficient
of the resistivity \cite{Tamura03ntc} and the origin of the pseudogap in the 
electronic density of states (DOS). \cite{Ishii01,Tamura04hybridization}

Theoretically the binary QCs require a more sophisticated approach than the
above-mentioned ternary QCs because of the relatively heavy elements they 
contain:
As we will show, the (relativistic) SO interaction and electron-electron 
interactions that are not described by the usual local density approximation 
(LDA), cannot be ignored.
Because these effects are most prominent in \textit{i}-CdYb, we will focus on
this QC.

\begin{figure}
  \includegraphics[width=8.0cm]{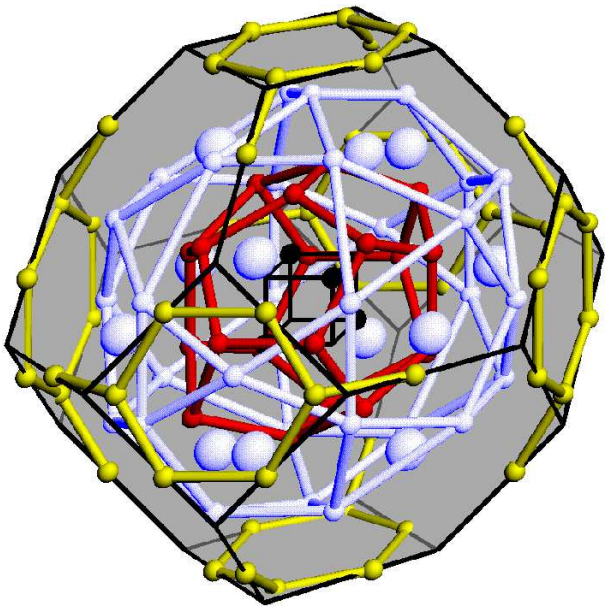}
  \caption{\label{fig_xyz}(Color online) Unit cell of 1/1 Cd$_6$Yb
  (spacegroup I23). 
  The Cd (small) and Yb atoms (large) form a shell structure.
  In the center Cd atoms are located on four vertices of a cube.
  The next shell (the dodecahedral cavity) consists of Cd atoms placed on the 
  vertices of a dodecahedron.
  The third shell is a Cd icosidodecahedron composed of triangles and 
  pentagons. 
  Yb atoms are located at the centers of the pentagons.
  The remaining Cd atoms and Cd atoms of the icosidodecahedra in neighboring 
  unit cells (not shown) form a defect triacontahedron 
  (Ref.\ \onlinecite{Gomez03}).}
\end{figure}

1/1 Cd$_6$Yb \cite{Palenzona71} is a rational approximant \cite{Takakura01} 
to \textit{i}-CdYb.
Its structure can be described \cite{Gomez03} as a bcc packing of 
interpenetrating icosahedral clusters (Fig.\ \ref{fig_xyz}).
\textit{i}-CdYb is composed of the same clusters. \cite{Ryan01,Sato02}
Whereas the exact structure of \textit{i}-CdYb has not been 
determined, the structure of 1/1 Cd$_6$Yb is known.
Some uncertainty exists only about the positions of the four Cd atoms inside 
the so-called dodecahedral cavity (Fig.\ \ref{fig_xyz}).
1/1 Cd$_6$Ca is isostructural with 1/1 Cd$_6$Yb: Ca atoms replace Yb.
In this paper we present an \textit{ab initio} electronic structure study
of 1/1 Cd$_6$Yb and 1/1 Cd$_6$Ca. 
Because of the structural proximity of the binary icosahedral QCs and their
1/1 approximants we expect that our results that do not depend on the 
long-range order of these materials, are also valid for \textit{i}-CdYb and 
\textit{i}-CdCa.

We studied 1/1 Cd$_6$Yb and 1/1 Cd$_6$Ca with the computer program 
\textsc{wien}2k, \cite{Schwarz02version3} an implementation of the 
full-potential linear augmented plane-wave method \cite{Singh94} based on 
density functional theory. \cite{Kohn65}
For the exchange and correlation potential we used the LDA. \cite{Perdew92}
In our DOS calculations we included the SO coupling using the so-called
``second variational treatment.'' \cite{MacDonald80}

Although the LDA gives reliable optimal atomic positions for many materials,
for the same materials it may fail to predict the proper electronic ground 
state. \cite{Czyzyk94}
In particular, the LDA description of the electronic structure of compounds
containing (strongly correlated) transition-metal \textit{d} or rare-earth
\textit{f} electrons is often inadequate. \cite{Anisimov97}
As we will show below,
a good example of this shortcoming of the LDA is the wrong prediction 
\cite{Ishii01} for the position of the Yb-4\textit{f} states in the DOS of 
1/1 Cd$_6$Yb.
A qualitative improvement is achieved by the LDA + U method, \cite{Anisimov93}
which takes into account the orbital dependence of the Coulomb and exchange 
interactions, which is absent in the LDA. \cite{Anisimov97}

We applied the LDA + U formalism of Ref.\ \onlinecite{Anisimov93} to the 
Cd-4\textit{d} and the Yb-4\textit{f} states.
The orbital potentials felt by these states depend on two adjustable 
parameters: the Coulomb parameter $U$ (also called Hubbard $U$) and the 
exchange parameter $J$.
For fully occupied orbitals like the Cd-4\textit{d} and the Yb-4\textit{f} 
states in 1/1 Cd$_6$Yb it is straightforward to show that the effect of $U$ 
and $J$ is to shift the orbital potential by $- \frac{1}{2} (U - J)$.
Therefore we could choose $J = 0$ without loss of generality.
We adjusted $U$ to find the best possible agreement with a photo-emission
(PE) spectrum of Ref.\ \onlinecite{Tamura02pes} and a PE spectrum that we 
present in this paper.
Our optimal values are $U$(Cd) $= 5.6$ and $U$(Yb) $= 3.1$ eV

\begin{figure}
  \includegraphics[width=3.6cm]{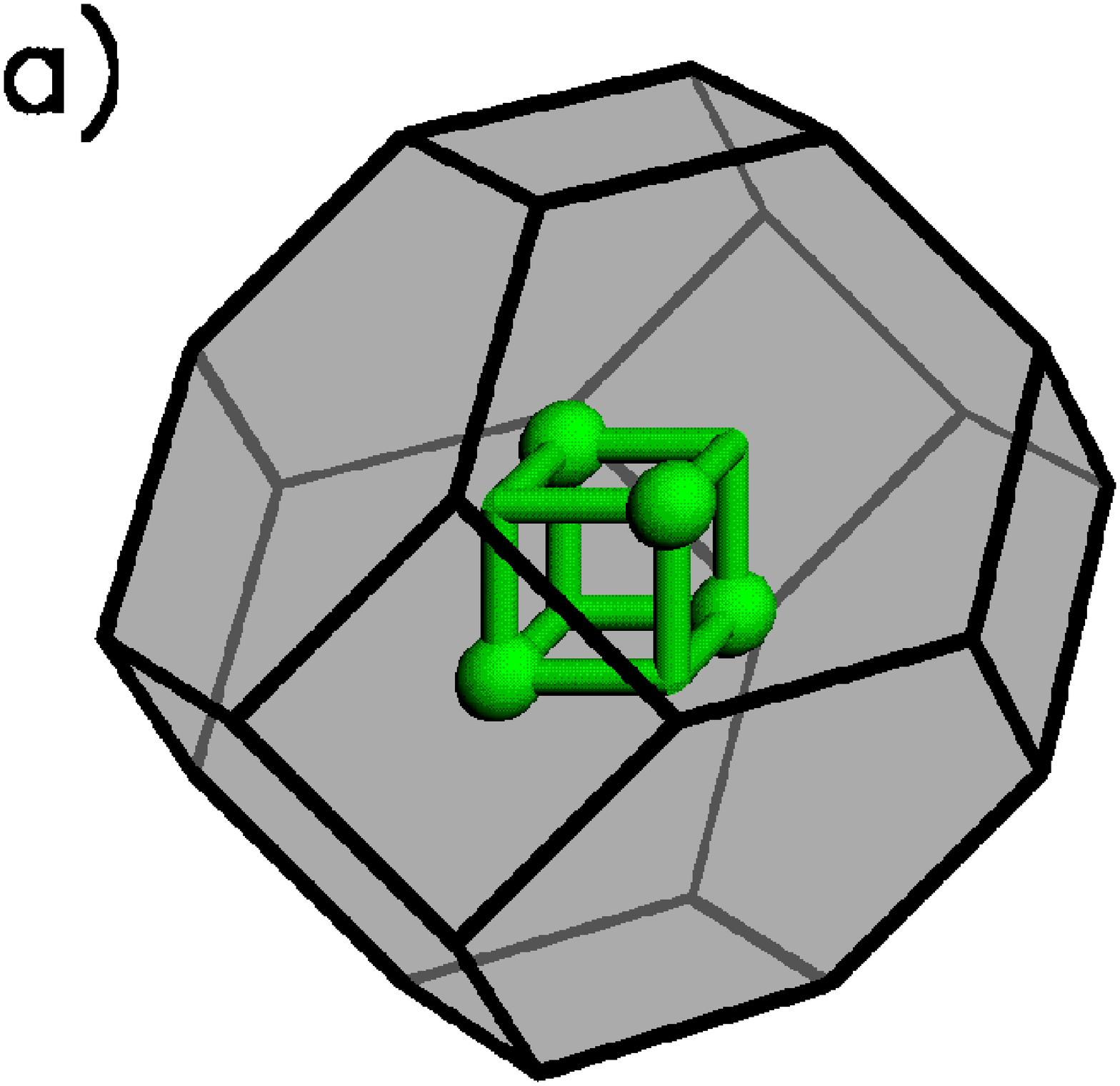}
  \includegraphics[width=3.6cm]{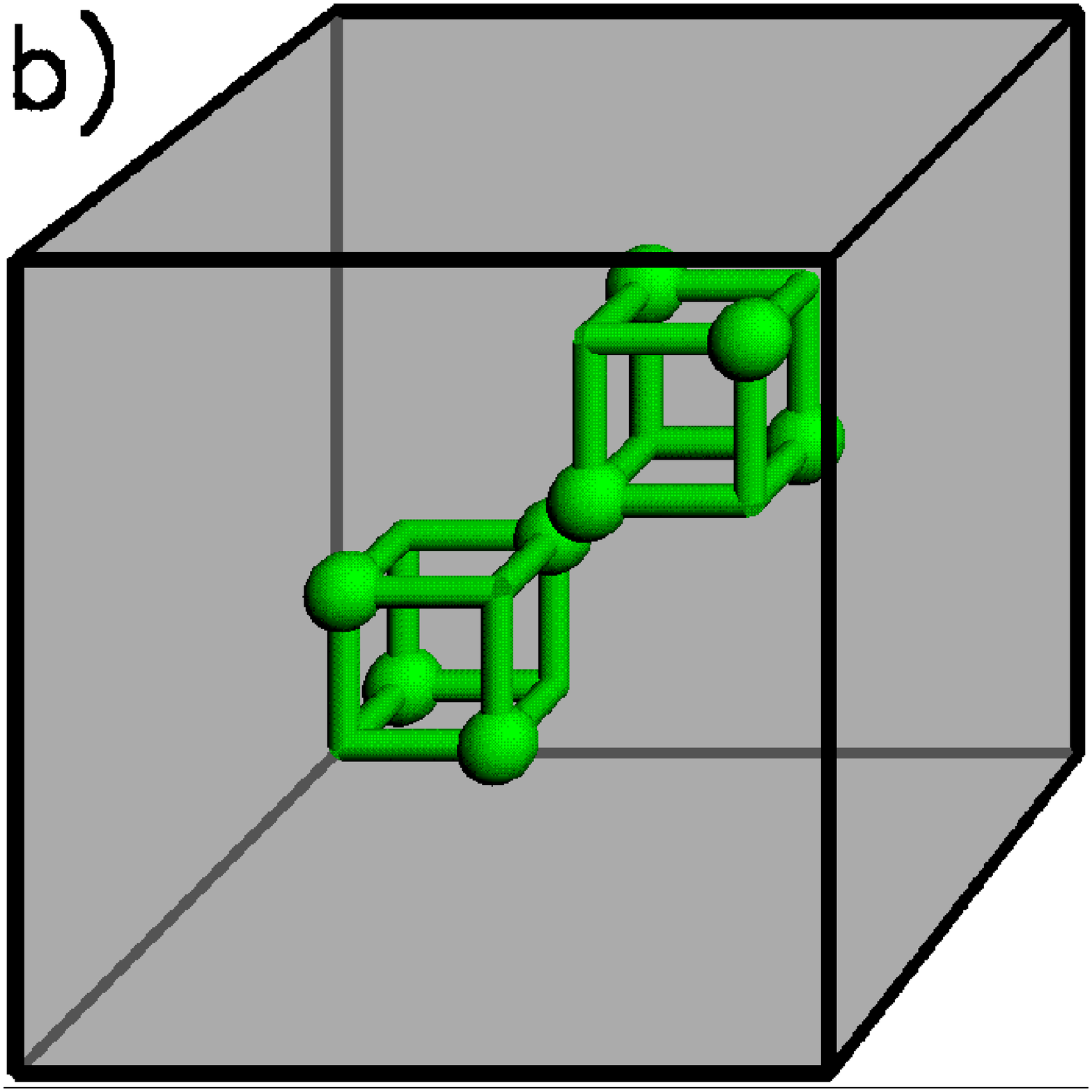}
  \includegraphics[width=3.6cm]{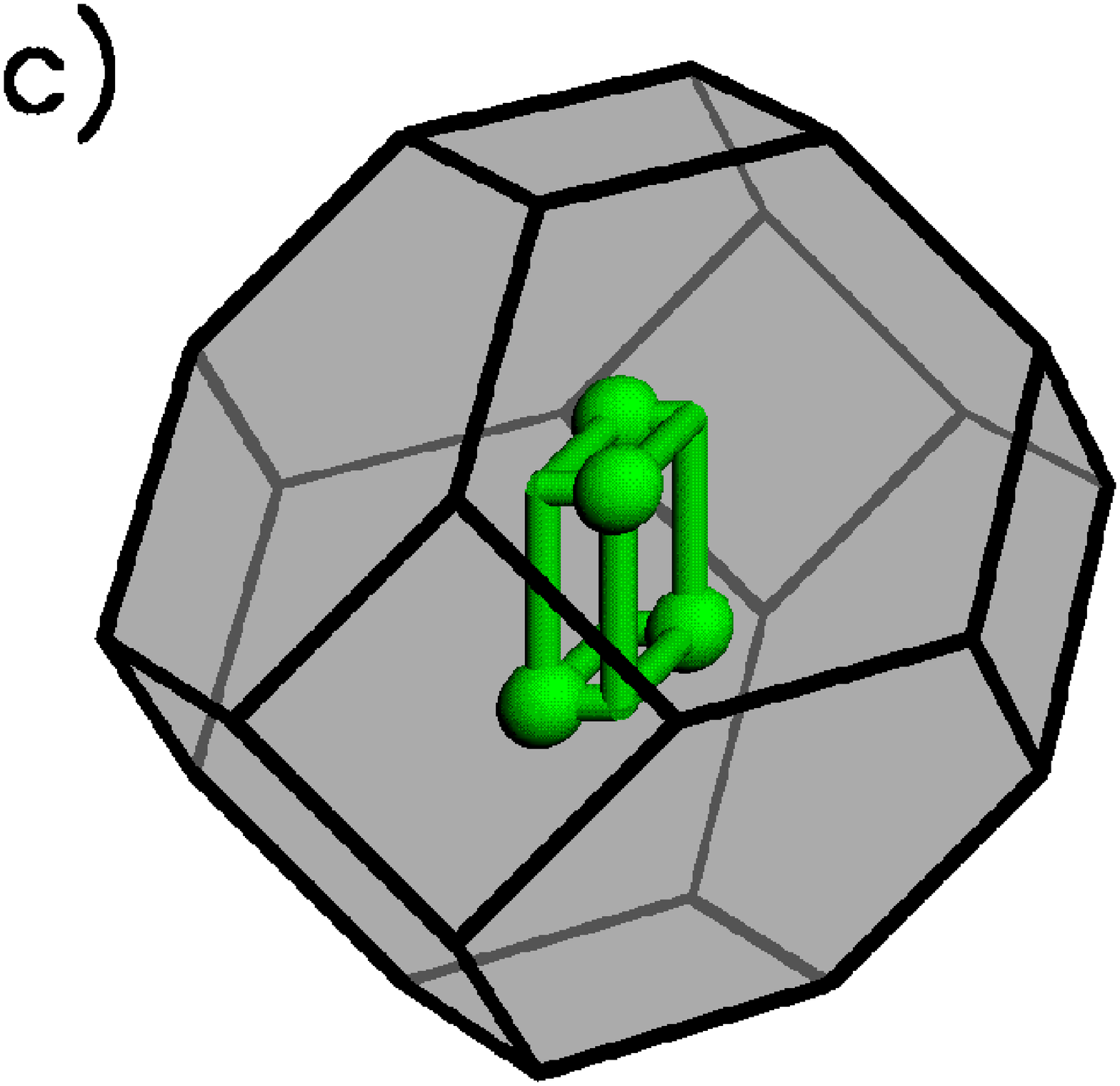}
  \includegraphics[width=3.6cm,clip]{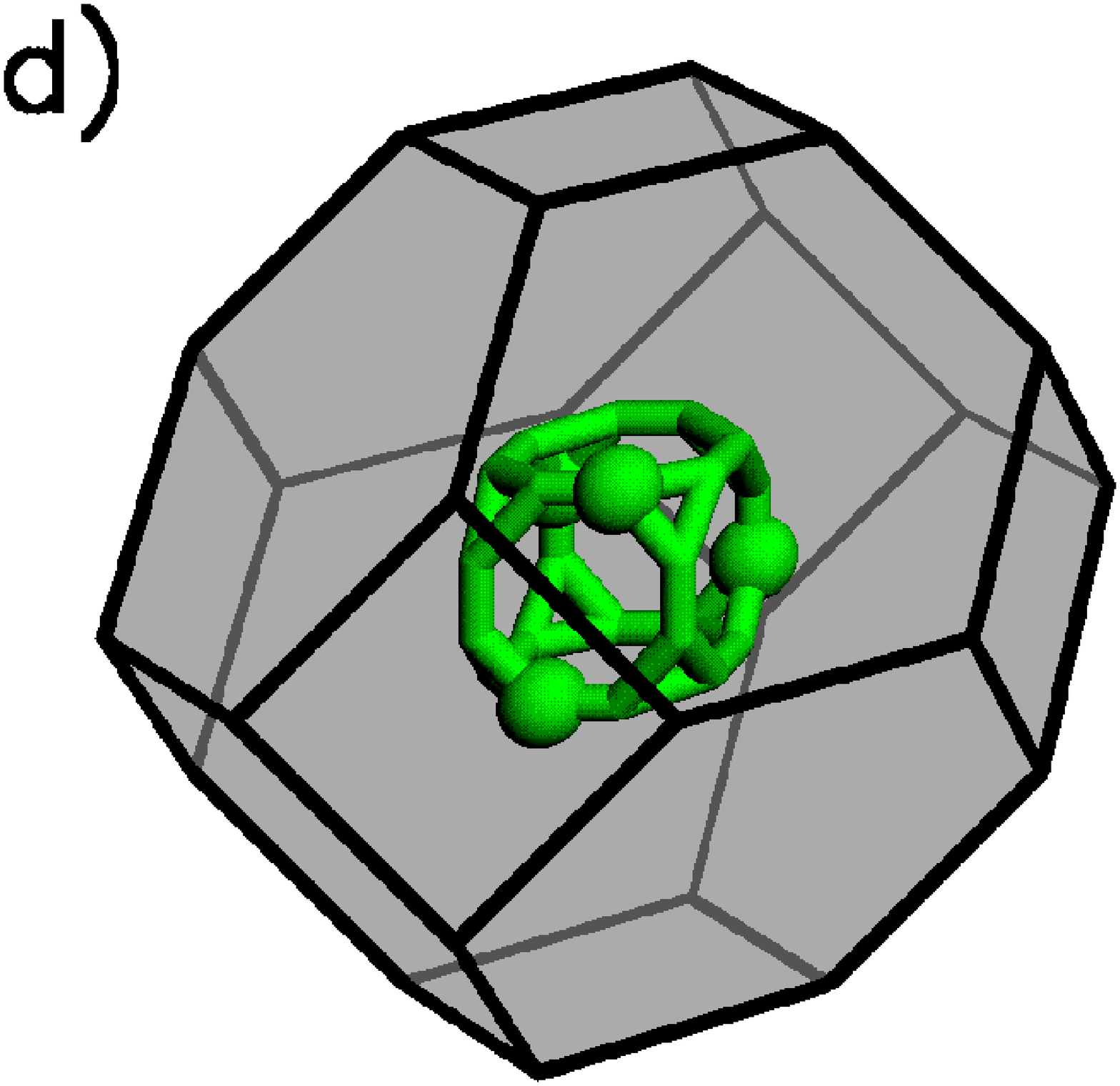}
  \includegraphics[width=3.6cm]{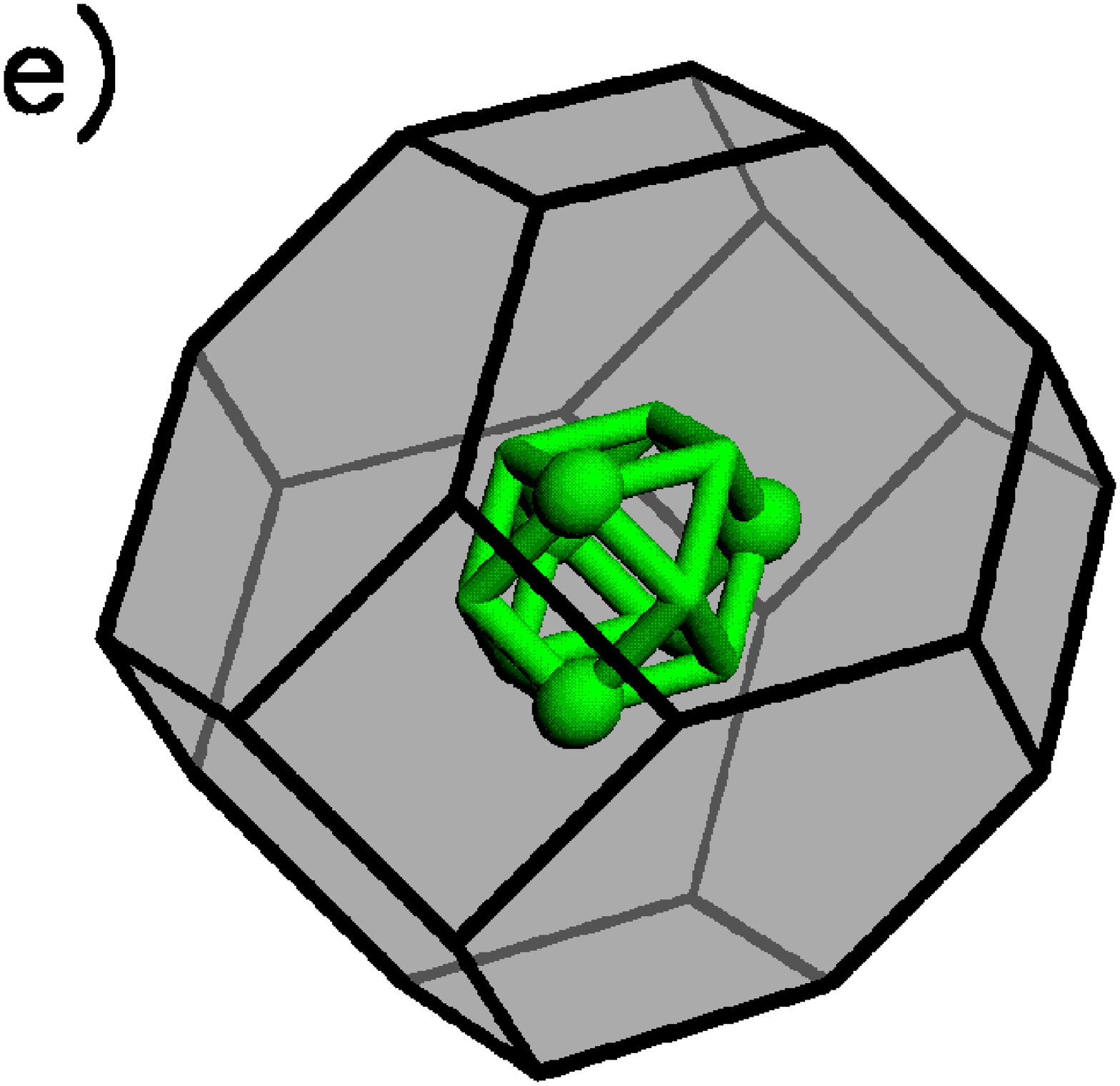}
  \caption{\label{fig_symmetry}(Color online) Models of 1/1 Cd$_6$Yb differ 
  from each other in the positions of the Cd atoms inside the dodecahedral
  cavity (Fig.\ \ref{fig_xyz}).
  These positions and the unit cells are schematically indicated for models 
  with symmetries (a) I23, (b) Pn$\bar{\textrm{3}}$, (c) I222, (d) C2, and
  (e) Imm2 (See text).}
\end{figure}

Different structural models have been proposed for 1/1 Cd$_6$Yb.
They differ essentially only in the positions of the four Cd atoms inside
the dodecahedral cavity (Fig.\ \ref{fig_xyz}).
Structural details are often important for the electronic structure.
Therefore we made a comparison of the various structural models.
In the original model \cite{Palenzona71} of 1/1 Cd$_6$Yb the Yb and Cd atoms 
are on the same positions (scaled by the lattice parameter $a$) as Ru and Be 
in Ru$_3$Be$_{17}$ (Ref.\ \onlinecite{Sands62}).
Four additional Cd atoms are located \cite{Palenzona71} on four vertices of a 
cube [Fig.\ \ref{fig_symmetry}(a)] inside the dodecahedral cavity.
The spacegroup of this original model is I23 (Ref.\ \onlinecite{Palenzona71}).

\begin{figure}
  \includegraphics[angle=270,width=8.0cm]{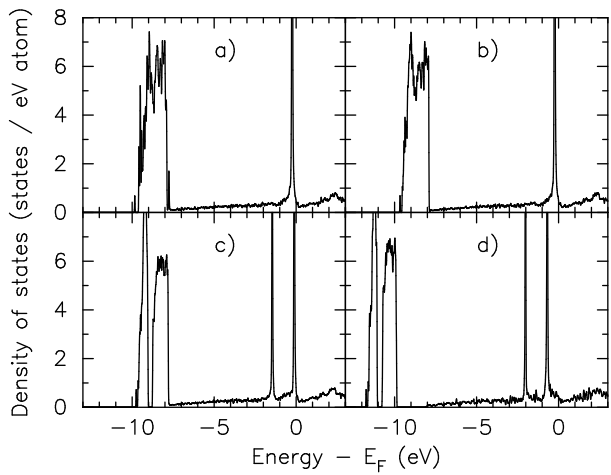}
  \caption{\label{fig_dos}Electronic DOS of 1/1 Cd$_6$Yb (spacegroup I23).
  (a) Original model (Ref.\ \onlinecite{Palenzona71}). 
  (b)--(d) Relaxed model (Ref.\ \onlinecite{Zijlstra04mrs}).
  In (c) and (d) the SO interaction is included.
  (d) was calculated with the LDA + U method.
  All DOSs were convoluted with a Gaussian with a full width at half maximum
  of $30$ meV.
  The DOS in (d) is not as smooth as the other DOSs, because fewer \textbf{k}
  points were used to calculate it.}
\end{figure}

To calculate a good electronic DOS it is in general necessary to use accurate 
atomic positions. \cite{Zijlstra03almnsi,Zijlstra04alcufe,Zijlstra04mrs}
In particular, idealized positions may lead to a DOS with unphysical features 
(spikes), \cite{Zijlstra03almnsi} a pseudogap at the Fermi energy $E_F$ that 
is too narrow, \cite{Zijlstra03almnsi} or shifted peaks. 
\cite{Zijlstra04alcufe}
Therefore we have performed \cite{Zijlstra04mrs} a structural relaxation of 
the atomic positions of the original model of 1/1 Cd$_6$Yb.
Our final positions \cite{Zijlstra04mrs} are relatively close to the original 
positions. \cite{Palenzona71}
During the structural relaxation the total energy decreased by $17$ meV / 
atom.
In the DOS the peak around $E_F - 9$ eV became $14$ \% narrower
[Figs.\ \ref{fig_dos}(a) and \ref{fig_dos}(b)],
indicating a reduced (spatial) overlap of the Cd-4\textit{d} states.

Use of the generalized gradient approximation \cite{Perdew96} (instead of the 
LDA) resulted in marginally different atomic positions.
The main changes in the DOS were a lowering of the Cd-4\textit{d} states by
$0.06$ eV and an increase of the Yb-4\textit{f} peak by $0.01$ eV.
In the present study we are not interested in such small differences.
Therefore, we present results for the LDA only.

Electrical resistivity, specific heat, and electron diffraction experiments
have suggested \cite{Tamura02pn3} that at low temperatures 
($T < 110$ K) a structure with a simple cubic unit cell (spacegroup 
Pn$\bar{3}$) is favored over the bcc I23 structure.
(Later experiments \cite{Tamura04noncubic} have indicated that the 
low-temperature phase is noncubic, but for computational reasons we will
consider noncubic structures only in the body-centered models.) 
In the Pn$\bar{3}$ model the Cd atoms inside the dodecahedral cavities occupy
vertices of cubes with an alternating orientation [Fig.\ 
\ref{fig_symmetry}(b)].
The DOSs of the original I23 and Pn$\bar{3}$ models are not significantly
different. \cite{Ishii02}
We relaxed the Pn$\bar{3}$ model.
The total energy of the relaxed Pn$\bar{3}$ model was 
$0.4$ meV / atom lower than that of the relaxed I23 model. 
Further convergence studies would be necessary to establish whether this 
number is significant. 
For the present study, however, it suffices to say that we found that the 
DOSs of the relaxed I23 and Pn$\bar{3}$ models were practically identical,
and that we therefore restricted ourselves to the computationally less 
involved body centered models, such as the I23 model [Fig.\ 
\ref{fig_symmetry}(a)].

Another modification of the I23 model has been proposed on the basis of a
recent x-ray diffraction study. \cite{Gomez03}
According to this study the central Cd atoms are located on vertices of a 
truncated cube [Fig.\ \ref{fig_symmetry}(d)].
When we tried to occupy four of these vertices, we found that there are 
only two inequivalent ways to do this (we discarded Cd-Cd nearest neighbor 
distances that were shorter than $2.4$ \AA).
The first possibility is that the Cd atoms are on vertices of a rectangular
parallelepiped [Fig.\ \ref{fig_symmetry}(c)].
This structure has spacegroup I222.
When we relaxed these atomic positions we found that the atoms moved to the
positions corresponding to the I23 symmetry [Fig.\ \ref{fig_symmetry}(a)],
indicating that the I222 structure is not stable.
The second possibility is indicated in Fig.\ \ref{fig_symmetry}(d).
The spacegroup is C2. 
When we relaxed the atomic coordinates of this model we found that the Cd 
atoms moved to positions somewhere in between vertices of a cube octahedron 
[Fig.\ \ref{fig_symmetry}(e)] and vertices of an icosahedron.
The spacegroup of this model is Imm2.
The Imm2 model had a total energy that was $2$ meV / atom lower than that of 
the relaxed I23 model, indicating that the Imm2 structure is more stable than 
the I23 structure.
The peak positions in the DOSs of the relaxed Imm2 and I23 models were equal 
within $0.01$ eV.

Summarizing our results so far, we found that the different models of 
1/1 Cd$_6$Yb had similar electronic DOSs.
The only significant change was a narrowing of the Cd-4\textit{d} peak after
the structural relaxation of the original I23 model.
It is of particular interest that
the main features in the DOS were found to be insensitive to the precise
positions of the four Cd atoms inside the dodecahedral cavity, which is 
consistent with the theoretical result of Ref.\ \onlinecite{Ishii02} that
these Cd atoms are weakly bound and do not play an important role in 
stabilizing 1/1 Cd$_6$Yb.
In the rest of this paper we present results for the model with the highest
symmetry, i.e., the I23 model.

We calculated the electronic DOSs of 1/1 Cd$_6$Yb and 1/1 Cd$_6$Ca.
A separate structural relaxation was performed for 1/1 Cd$_6$Ca.
Figure \ref{fig_dos}
shows our results for 1/1 Cd$_6$Yb obtained at various levels of 
sophistication.
The result of Fig.\ \ref{fig_dos}(a) was obtained without SO coupling or
orbital potentials (Hubbard $U$s) for the original I23 model.
It is essentially equal to the result of Ref.\ \onlinecite{Ishii01}.
The main features are the Cd-4\textit{d} peak around $E_F - 9$ eV, one
Yb-4\textit{f} peak just below the Fermi energy, and a pseudogap near $E_F$.
In Ref.\ \onlinecite{Ishii01} it has been shown that this stabilizing 
pseudogap is formed as a consequence of 
hybridization between the free-electronlike \textit{s} and \textit{p} states
and low-lying (unoccupied) Yb-5\textit{d} states.

The DOSs of Ref.\ \onlinecite{Ishii01} and Fig.\ \ref{fig_dos}(a) are in 
disagreement with a PE spectrum \cite{Tamura02pes} of 1/1 Cd$_6$Yb, in which 
there are two Yb-4\textit{f} peaks around $2.0$ and $0.7$ eV below the Fermi 
energy.
The $1.3$ eV split between these peaks is attributed \cite{Tamura02pes} to
the SO interaction.
It has been suggested \cite{Tamura02pes} that differences between the 
Yb-4\textit{f} peak positions in the DOS of Ref.\ \onlinecite{Ishii01} and in
the PE spectrum of Ref.\ \onlinecite{Tamura02pes} are due to a shape 
approximation for the potential that has been made in Ref.\ 
\onlinecite{Ishii01}.
The fact that our full-potential calculation is essentially equal to the
result of Ref.\ \onlinecite{Ishii01} shows that this is wrong.
In fact, a comparison of Figs.\ \ref{fig_dos}(b) and \ref{fig_dos}(c)
shows that the most obvious discrepancy between Fig.\ \ref{fig_dos}(a) and 
the experiment, \cite{Tamura02pes} namely,
the absence of the SO splitting in Fig.\ \ref{fig_dos}(a), is a result of the 
absence of the SO interaction in the calculation.

\begin{figure}
  \includegraphics[angle=270,width=8.0cm]{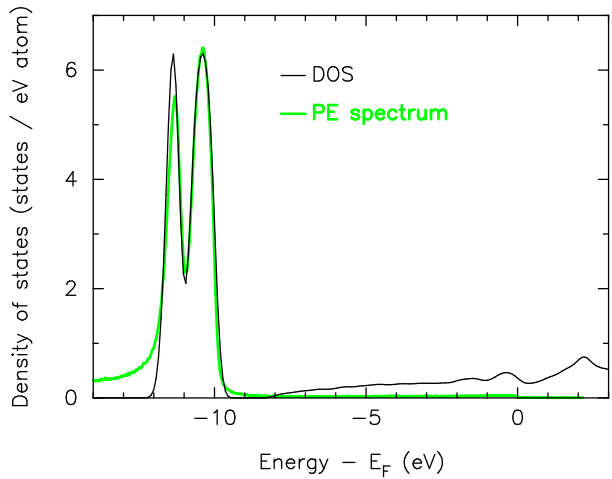}
  \caption{\label{fig_pes_cdca}(Color online) Comparison of the calculated DOS
  of 1/1 Cd$_6$Ca with a PE spectrum (See text). 
  The DOS was convoluted with a Gaussian with a full width at
  half maximum of $400$ meV.}
\end{figure}

As mentioned above the structural relaxation of the I23 model of 1/1 Cd$_6$Yb
led to a narrowing of the Cd-4\textit{d} peak [Figs.\ \ref{fig_dos}(a) and
\ref{fig_dos}(b)].
The SO interaction split the Yb-4\textit{f} and Cd-4\textit{d} peaks [Fig.\
\ref{fig_dos}(c)].
By including a Hubbard-$U$ term in the Hamiltonian for the Cd-4\textit{d} and
Yb-4\textit{f} states we could shift these peaks to lower energies [Fig.\
\ref{fig_dos}(d)].
We adjusted the $U$s to get the best possible fit with the PE spectra shown
in Figs.\ \ref{fig_pes_cdyb} and \ref{fig_pes_cdca}.
\begin{figure}
  \includegraphics[angle=270,width=8.0cm]{pes_cdyb}
  \caption{\label{fig_pes_cdyb}(Color online) Comparison of the calculated DOS
  of 1/1 Cd$_6$Yb with a PE spectrum (Ref.\ \onlinecite{Tamura02pes}).
  The DOS was convoluted with a Gaussian with a full width at
  half maximum of $200$ meV.}
\end{figure}
Our final DOSs reproduced the experimental peak positions very well
(Figs.\ \ref{fig_pes_cdyb} and \ref{fig_pes_cdca}). 

The PE spectrum of 1/1 Cd$_6$Ca (Fig.\ \ref{fig_pes_cdca}) was measured by 
x-ray PE spectroscopy with an incident photon energy of $350$ eV at the 
temperature of $20$ K using a GAMMADATA-SICENTA SES2002 hemispherical 
analyzer. 
Details of the sample preparation have been given elsewhere.
\cite{Tamura04hybridization}
The Cd-4\textit{d} states are greatly enhanced in the experimental spectrum 
since at this photon energy the cross section of the Cd-4\textit{d} states is 
the largest among all the states that may contribute in this binding energy 
range: 
For instance, the cross section of Cd-4\textit{d} is one or two orders of 
magnitude larger than those of Cd-5\textit{s} and Ca-4\textit{s}.

The SO interaction and the strong electronic correlations in 1/1 Cd$_6$Yb
are not only important for the peak structure in the DOS, they are also
essential for a proper description of the DOS near the Fermi energy.
This is illustrated by a comparison of the electronic contributions to the 
specific heat,
\begin{equation}
\label{eq_gamma}
\gamma = \frac{\pi^2}{3} k_B^2 \mathrm{DOS}(E_F) (1 + \lambda),
\end{equation}
calculated at various levels of sophistication (Table \ref{table_gamma}),
\begin{table}
  \caption{\label{table_gamma}
  Calculated and measured electronic contributions to the specific heat.
  The values are given per ``mole formula unit'', viz., Cd$_6$Yb and 
  Cd$_{5.7}$Yb for the 1/1 approximant and the QC, respectively.}
  \begin{ruledtabular}
  \begin{tabular}{ldd}
  Calculation / Experiment 
   & \multicolumn{2}{c}{$\gamma$ (mJ mol$^{-1}$ K$^{-2}$)} \\
   & \multicolumn{1}{c}{1/1 Cd$_6$Yb} & \multicolumn{1}{c}{\textit{i}-CdYb} \\
  \hline
  LDA calculation [Fig.\ \ref{fig_dos}(a)]          &  7.3        \\
  LDA calculation [Fig.\ \ref{fig_dos}(b)]          &  7.4        \\
  LDA + SO calculation [Fig.\ \ref{fig_dos}(c)]     & 11.8        \\
  LDA + SO + U calculation [Fig.\ \ref{fig_dos}(d)] &  6.1        \\
  Experiment (Ref.\ \onlinecite{Tamura01transport}) &      & 19.2 \\
  Experiment (Ref.\ \onlinecite{Pope01})            &      &  7   \\
  Experiment (Ref.\ \onlinecite{Dhar02})            & 51   &  7.5 \\
  \end{tabular}
  \end{ruledtabular}
\end{table}
where we assumed that the phonon-correction factor $\lambda = 0$.
(For a proper comparison with experiment $\lambda$ can, of course, not be 
ignored. \cite{Tamura01transport})
Experimental values of $\gamma$ are also given (Table \ref{table_gamma}).
It is remarkable \cite{Dhar02} that $\gamma$ is much lower in the QC
than in the 1/1 approximant (Table  \ref{table_gamma}).
This indicates a substantial rearrangement of states leading to a considerable
lowering \cite{Dhar02} of the DOS at the Fermi energy in \textit{i}-CdYb.
This rearrangement of states could be due purely to geometrical
effects (Hume-Rothery), or geometrical effects coupled with increased
hybridization between the Yb-5\textit{d} states and the free-electronlike 
\textit{s} and \textit{p} states, a scenario that still remains to be 
explored. 

In 1/1 Cd$_6$Yb the Yb-4\textit{f} states are fully \cite{Tamura01transport} 
occupied.
The states at the Fermi energy have essentially no Yb-4\textit{f} character.
\cite{Tamura01transport}
Therefore, the Yb-4\textit{f} states do not contribute \cite{Tamura01transport}
to transport properties or to the temperature dependence of thermodynamic 
properties.
This is in clear contrast to the mixed-valent compound YbCuAl, in which the 
Yb-4\textit{f} electrons are important for the heavy fermion character. 
\cite{Stewart84}

In conclusion, the four Cd atoms inside the dodecahedral cavity occupied 
vertices of a (deformed) cube octahedron in the most stable model of 1/1 
Cd$_6$Yb.
The main features in the electronic DOS were insensitive to the model used.
To get a good agreement with PE spectra it was necessary to include the SO 
interaction in the Hamiltonian, which is responsible for the splitting of the 
Cd-4\textit{d} and the Yb-4\textit{f} peaks in the DOS.
In addition, the Cd-4\textit{d} and Yb-4\textit{f} electrons have strong 
correlations that cannot be treated in the LDA.
We could describe these correlations with the LDA + U method by including a
Hubbard-$U$ term in the Hamiltonian for the Cd-4\textit{d} and the 
Yb-4\textit{f} states. 

Financial support for this work was provided by the Natural Sciences and
Engineering Research Council of Canada.
Part of our calculations were performed on computers of SHARCNET (Hamilton,
Canada).
R.\ T.\ is indebted to Dr.\ T.\ Takeuchi and Prof.\ S.\ Takeuchi for a
collaboration in the experiment.


\end{document}